\documentstyle[aps,epsfig,graphicx,floats]{revtex}                              

\begin{document}

\newcommand{\amodel}{$\alpha$-model}
\newcommand{\aMODEL}{$\alpha$-MODEL}
\newcommand{\nctlattice}{$nct$-lattice}
\newcommand{\nctLATTICE}{$nct$-LATTICE}
\newcommand{\qmodel}{$q$-model}
\newcommand{\qMODEL}{$q$-MODEL}
\newcommand{\None}{N_{1}} 
\newcommand{\Ntwo}{N_{2}}
\newcommand{\none}{n_{1}} 
\newcommand{\ntwo}{n_{2}}
\newcommand{\Tone}{T_{1}} 
\newcommand{\Ttwo}{T_{2}}
\newcommand{\tone}{t_{1}} 
\newcommand{\ttwo}{t_{2}}
\newcommand{\Cone}{C_{1}} 
\newcommand{\Ctwo}{C_{2}}
\newcommand{\cone}{c_{1}}
\newcommand{\ctwo}{c_{2}}
\newcommand{\azero}{\alpha_{0}} 
\newcommand{\aone}{\alpha_{1}} 
\newcommand{\atwo}{\alpha_{2}}
\newcommand{\eonev}{{\bf e}_1}
\newcommand{\etwov}{{\bf e}_2} 
\newcommand{\rv}{{\bf r}}
\newcommand{\gv}{{\bf g}} 
\newcommand{\xv}{{\bf x}} 
\newcommand{\yv}{{\bf y}}
\newcommand{\zv}{{\bf z}} 
\newcommand{\sm}{{\bf \sigma}}

\title{Average stresses and force fluctuations in non-cohesive granular 
  materials}
\author{Joshua E. S. Socolar} \address{Department of Physics and
  Center for Nonlinear and Complex Systems, Duke University, Durham,
  NC 27708} 
\date{\today}

\maketitle

\begin{abstract}
  A lattice model is presented for investigating the fluctuations in
  static granular materials under gravitationally induced stress.
  The model is similar in spirit to the scalar \qmodel\ of 
  Coppersmith {\em et al.}, but
  ensures balance of all components of forces and torques at each site.
  The geometric randomness in real granular materials is modeled
  by choosing random variables at each site, 
  consistent with the assumption of cohesionless grains.
  Configurations of the model can be generated rapidly, allowing the 
  statistical study of relatively large systems.  
  For a 2D system with rough walls, the model generates configurations
  consistent with continuum theories for the average stresses
  (unlike the \qmodel) without requiring the assumption of
  a constitutive relation.
  For a 2D system with periodic boundary conditions, the model
  generates single-grain force distributions similar to those
  obtained from the \qmodel\ with a singular distribution of $q$'s.
\end{abstract}


The microscopic stress field in a static granular material has an 
extraordinarily complex structure.
Viewed from the perspective of standard elasticity theory,
the geometric disorder in the packing of grains gives rise
to extremely complicated boundary conditions on the stress
equilibrium equations.
In general, this disorder is rather difficult to characterize
statistically and may even exhibit nontrivial correlations
induced by the dynamical history of the material.

At present, the overwhelming majority of attempts to model
granular materials are formulated at the level of 
a continuous stress field, which is intended to represent
a smoothed version of actual stresses, averaged over regions
large compared to the grain size.
In order to close the system of stress equilibrium equations,
a constitutive relation must be assumed, such as the 
Mohr-Coulomb condition that the material is on the verge
of yielding at everywhere within a plastic zone.
While this condition \cite{nedderman}
and others like it \cite{wittmer,cantelaubePG}
have met with some success in many different situations, 
they rest ultimately on ad hoc
assumptions about the connection between the 
microscopic and macroscopic stresses.

One factor that could in principle pose fundamental difficulties
for continuum theories for the average stress is
that fluctuations in the microscopic stresses may be quite large, 
perhaps large enough to invalidate typical assumptions about the scales
over which the material can be modeled as a continuum. 
In several recent experiments that directly image the stress field,
stress chains (filamentary configurations of strongly stressed grains)
have been observed with correlation lengths that are apparently
limited only by the system size 
(though the systems have not been much larger than 30 grain diameters 
in the relevant dimension). \cite{nagel,howellPG}
For these reasons it is important to obtain some theoretical understanding of
what determines the size and spatial structure of the fluctuations.

Coppersmith {\it et al.} recently stimulated interest in a
simplified statistical approach to stresses in granular materials with
the introduction of the ``\qmodel'', a lattice model whose
configurations are intended to represent the way in which vertical
forces are supported in a non-cohesive material.
In such a model it is possible for very large fluctuations to arise,
even on scales as small as a lattice constant.
The constitutive assumptions of continuum models are replaced here
by an ansatz concerning the form of the microscopic effects of 
geometric disorder in the material.
The central result of Ref.~\cite{qmodel1} is that for infinitely wide layers 
(or materials subject to periodic boundary conditions 
in the horizontal directions),
fluctuations in the vertical forces supported by grains at a given
depth are of the same magnitude as the average force supported at that depth.  
An elegant calculation shows that the probability distribution
for the vertical force supported by a grain deep in the pile has an
exponential tail, rather than Gaussian. \cite{qmodel1}

For all its merits, the \qmodel\ has three serious flaws.  
First, it takes no account of the constraints imposed by 
horizontal force balance and torque balance on each grain, 
and thus does not contain the proper conservation laws at the microscopic level.  
This may constitute a flaw that leads to incorrect predictions,
though it is also possible
that the constraints in question do not affect the large scale behavior.
Second, when studied in the
silo geometry, the \qmodel\ yields predictions for the average stresses
that dramatically disagree with classical theories and experiments.
\cite{pfPG}
While quantitative agreement may not be expected given the
crude representation of the boundary conditions that must be used
in constructing \qmodel\ configurations,
the qualitative discrepancy is striking, as explained below.
Finally, there is no clear procedure for connecting the lattice
constant in the \qmodel\ to a physical length scale.

In this paper, a new model is presented that explicitly incorporates the 
the relevant force and torque balance constraints into the
lattice approach of the \qmodel\ and provides a natural connection
between the lattice constant and the grain size.
Stress distributions are then computed for the silo geometry with 
force-bearing walls and with periodic boundary conditions. 
It is shown that the new model gives much better
agreement with previous theories and experiments in the silo geometry,
and thus appears to be a more reliable basis for investigating the
subtleties of the stress fluctuations.  
Numerical results are then presented  for the single grain
weight distribution from the new model at large depths.
The single-grain weight distributions are similar in form to
those obtained in the \qmodel\ and therefore may be thought of
as providing a firmer foundation for the \qmodel\ predictions.

It is useful to make a conceptual distinction
between the definition of the basic lattice with 
appropriate variables defined on it
and the assumptions about those variables that are relevant for
the study of non-cohesive granular materials.
The geometric structure of the model is a square lattice with
variables representing net normal forces, couples, and tangential forces
on each edge.
This structure will be referred to as the ``\nctlattice''.
Every possible stress field in any type of static medium can be
mapped to a configuration on the \nctlattice.  
(The mapping is many-to-one.)
In order to model a non-cohesive granular material,
several restrictions must be made on the values of the normal forces,
couples, and tangential forces at each edge and an ansatz must be made for
form of the disorder in the system.
A particular model that incorporates these restrictions
will be called the ``\amodel'', 
for reasons that will become apparent below.

The paper is organized as follows: In Section~\ref{sec:amodel} the
\amodel\ is defined for 2D systems and the connections of the model
parameters to physical parameters and constraints are discussed.  
In Section~\ref{sec:silo} results are presented for the case of narrow silos
with rough walls in 2D and contrasted with results obtained from the \qmodel.
In Section~\ref{sec:deepbox} results are presented
for the case of wide 2D systems with periodic boundary conditions.
Section~\ref{sec:conclusions} includes a discussion of some general issues
and the generalization of the \amodel\ to three dimensions.

\section{The \aMODEL\ in two dimensions}\label{sec:amodel}

\subsection{The \nctlattice}
\begin{figure}[tbp]
  \centerline{
    \includegraphics[width=2.5in]{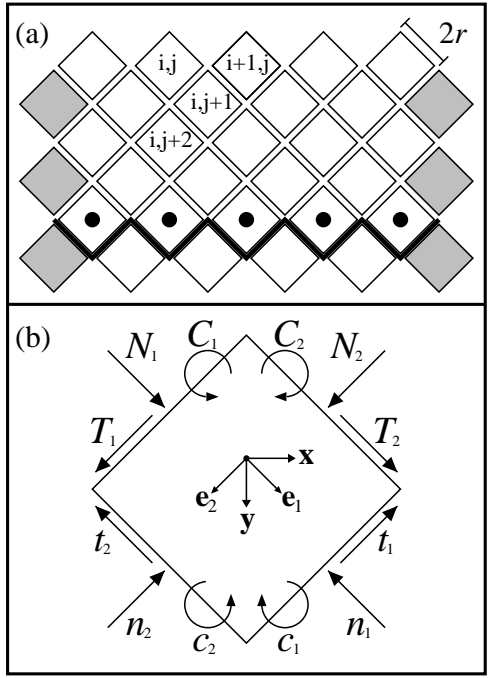}
  }
  \vspace{0.2in}
  \caption{Definition of the \amodel.
        (a) The lattice of square cells.  Shaded cells represent walls
        for the silo geometry and identified cells for 
        periodic boundary conditions.
        Dashed edges are assumed to transmit no force in the silo geometry.
        The cells marked with a large dot constitute a single layer.
        The thick edges are used to compute the force on a layer to compare
        with the Janssen solution. 
        (b) The variables used to describe the stress in a single cell
        and the unit vectors defining directions mentioned in the text.}
  \label{fig:amodel}
\end{figure}

Consider an arbitrary (possibly highly inhomogeneous) material that is
known to be in static stress equilibrium.  
For present purposes, the
material is taken to be two-dimensional and the following lattice
representation of the stress field is defined.  
(See Figure~\ref{fig:amodel}.)
A square lattice is constructed, with each cell representing 
a portion of the material.  
The length of each cell edge is $2r$.
A horizontal row of cells sharing vertices, such as those
marked with a dot in Figure~\ref{fig:amodel}, will be
called a ``layer'' of cells.
Cell $ij$ is defined as the $i^{th}$ cell from the left
in the $j^{th}$ layer from the top, with the left edge
being defined as shown in Figure~\ref{fig:amodel} for
even or odd $j$.
Note that $j$ increases downward, which is also defined
as the positive $\yv$ direction.

Associated with each edge in the lattice are a normal force, a
tangential force, and a couple.  
Taken together, these three quantities fully determine 
the force and torque exerted on one cell
about its center by the other cell sharing that edge.  
The couple is necessary to represent the torque that can be applied 
to a cell even in the absence of tangential forces, 
due to the manner in which the normal force is distributed along the edge.  
The couple is here defined so that it does {\em not} include 
the contribution of the tangential force to the torque.
The contribution from tangential forces is just equal to 
$r$ times the net amplitude of the tangential force, 
regardless of how that force is distributed along the edge.

Denote the amplitudes of the normal forces on the four edges of cell $ij$
by $\None^{(i,j)}$, $\Ntwo^{(i,j)}$, $\none^{(i,j)}$, $\ntwo^{(i,j)}$; 
the couples 
by $\Cone^{(i,j)} r$, $\Ctwo^{(i,j)} r$, $\cone^{(i,j)} r$, $\ctwo^{(i,j)} r$; 
and the tangential force amplitudes 
by $\Tone^{(i,j)}$, $\Ttwo^{(i,j)}$, $\tone^{(i,j)}$, $\ttwo^{(i,j)}$; 
as shown in Figure~\ref{fig:amodel}b.  
We will drop the superscript whenever this leads to no ambiguity.
Both $N$ and $n$ refer to
the amplitude of the net compressive force 
between the cells sharing an edge --
negative values would indicate a net tensile force between them.
The sign convention for tangential forces is chosen such that
positive $T$ or $t$ always indicates a positive downward component
of the force exerted 
{\em by} the cell with a higher center 
{\em on} the lower one. 
The sign convention for the couples is that
the directions indicated in Figure~\ref{fig:amodel}b 
correspond to positive values of each of the $c_i$ and $C_i$'s.
This choice allows the right-left symmetry of the model
to be immediately evident in the equations below.
Note that the uppercase variables associated with a given
cell are identically equal to the lowercase variables for
the cell sharing the relevant edge.
For example, $\None^{(i,j)} = \none^{(i,j-1)}$ (or $\none^{(i-1,j-1)}$) 
for $j$ even (or odd).

In terms of the canonically defined \cite{LLelasticity} 
microscopic stress tensor $\sm (\rv)$, 
these forces and couples can be written as integrals over 
their respective edges.
For example, letting $\eonev$ and 
$\etwov$ be unit vectors in the
$\yv-\xv$ and $\yv+\xv$ directions 
as shown in Figure~\ref{fig:amodel}b, 
and letting $s$ run from $-r$ to $r$ along the edge:
\begin{eqnarray}
  \None & = & -\int_{-r}^{r} ds \; \eonev \sm (s) \eonev , \label{eq:ndef} \\
  \Tone & = & -\int_{-r}^{r} ds \; \etwov \sm (s) \eonev , \label{eq:tdef} \\
  \Cone & = & -\frac{1}{r}\int_{-r}^{r} ds \; s\; \eonev \sm (s) \eonev ,
                                                           \label{eq:cdef} 
\end{eqnarray}
where the integrals are over surfaces (lines in 2D) that may cut through
grains and/or contacts between grains.

Static equilibrium at the microscopic level requires \cite{LLelasticity}
\begin{equation}
  \label{eq:stresseqV}
  \partial_k \sigma_{ik} - \rho g_i = 0,
\end{equation}
where $\rho$ is the local density and $\gv$ is the gravitational acceleration.
Requiring that the total force on a cell vanish, one finds:
\begin{equation}
  \label{eq:stresseq}
  \oint\sigma_{ik} dS_k = m g_i,
\end{equation}
where $m = \int\rho dV$ is the mass of the material in the cell and
$dS_k$ is the $k$th component of the outward normal to 
surface element $S$.
Taking $\gv$ to be in the positive $\yv$ direction (downward)
and performing the integral over the entire cell,
the $x$ and $y$ components of Eq.~(\ref{eq:stresseq} yield
the following equations for vertical and horizontal force balance:
\begin{eqnarray} 
  \label{eq:vforcebalance}
  \none + \tone + \ntwo + \ttwo & = & 
                        \None + \Tone + \Ntwo + \Ttwo + \sqrt{2} mg , \\
  \label{eq:hforcebalance}
  \none - \tone - \ntwo + \ttwo & = & \None - \Tone - \Ntwo + \Ttwo .
\end{eqnarray}
Similarly, the vanishing of the total torque 
about the center of the cell requires
\begin{equation}  
  \label{eq:momenteq}
  \oint dS_y \; x\;(\sigma_{xy} - \sigma_{yy})
  + \oint dS_x \; y\;(\sigma_{yx} - \sigma_{xx})
  - \int \rho g x dV = 0,
\end{equation}
which leads to
\begin{equation}  
  \label{eq:torquebalance}
  \tone - \cone - \ttwo + \ctwo = -\Tone - \Cone + \Ttwo + \Ctwo + 2 u,
\end{equation}
where $2 u = (\sqrt{2}/r)\int dV\;\rho g x$. 
(A torque about the center of the cell may be
exerted by gravity because the center of mass of the 
material in the cell need not coincide with the geometric center.)  

The lattice and variables just described, together with the fundamental
physical force and torque balance constraints, constitute the \nctlattice.
A ``configuration'' of the lattice denotes a set of values of
$\none^{(i,j)}$, $\ntwo^{(i,j)}$ 
$\cone^{(i,j)}$, $\ctwo^{(i,j)}$, 
$\tone^{(i,j)}$, $\ttwo^{(i,j)}$, 
$m^{(i,j)}$, and $u^{(i,j)}$ for all $ij$
that satisfies 
Eqs.~(\ref{eq:vforcebalance}), (\ref{eq:hforcebalance}), 
and (\ref{eq:torquebalance})
for every $ij$.
Each configuration is a discrete representation of a possible
stress field, and any possible stress field induces a configuration.
Thus the \nctlattice\ structure may be a useful tool for investigating
stresses in a wide range of inhomogeneous materials.

\subsection{Definition of the \amodel}

We now turn to the modeling of non-cohesive granular materials on
the \nctlattice.
In the spirit of the \qmodel\, it is assumed
that a typical stress distribution can be obtained 
by propagating the forces from top to bottom on the lattice.
This assumption is made for the sole purpose of allowing
the rapid construction of plausible force configurations.
In the \amodel, configurations are generated by descending
through the lattice, solving
Eqs.~(\ref{eq:vforcebalancea})-(\ref{eq:torquebalancea2}) for each cell.
The process begins by specifying
values of $m$ and $u$ for each cell and 
$\None$, $\Ntwo$, $\Tone$, $\Ttwo$, $\Cone$, and $\Ctwo$
for the cells in the top layer.

Given the values of the uppercase variables at a particular cell,
the six lowercase variables must be determined.
Since these variables are constrained by
Eqs.~(\ref{eq:vforcebalance}),~(\ref{eq:hforcebalance}), 
and~(\ref{eq:torquebalance}),
the space of possible solutions is then three dimensional
and can be parameterized by three real numbers $\azero$, $\aone$, and $\atwo$.
The manner in which these three numbers are determined at each cell
must reflect the physics of the material being modeled.

It is convenient to make the definitions 
$\cone\equiv\aone\;\none$,
$\ctwo\equiv\atwo\;\ntwo$, and 
$\azero\equiv \Ttwo + \Ctwo - \tone + \cone + u$.
The force and torque balance equations for a single cell can then be written as
\begin{eqnarray} 
  \label{eq:vforcebalancea}
  \none + \tone + \ntwo + \ttwo & = & 
                              \None + \Tone + \Ntwo + \Ttwo + \sqrt{2} mg , \\
  \label{eq:hforcebalancea}
  \none - \tone - \ntwo + \ttwo & = & 
                              \None - \Tone - \Ntwo + \Ttwo ,               \\
  \label{eq:torquebalancea1}
  \tone - \aone\none + \azero & = & \Ttwo + \Ctwo + u,  \\
  \label{eq:torquebalancea2}
  \ttwo - \atwo\ntwo + \azero & = & \Tone + \Cone - u,
\end{eqnarray}
where the torque equation has been split into two so that it may
be expressed in a symmetric form.
If $m$ and $u$ are given for every cell, then the microscopic
true stress field induces a unique configuration of 
triples $(\azero,\aone,\atwo)$ on the lattice.

In a non-cohesive granular material consisting of convex grains,
the grains can ``push'' on each other, but never ``pull'';
i.e., there can be no tensile forces in the material 
on scales larger than the grain size.  
This feature is incorporated into the \amodel\ by imposing two restrictions:
\begin{enumerate}
\item All normal force amplitudes $\none$ and $\ntwo$ must be positive;
\item All $\aone$'s and $\atwo$'s must lie in the interval $(-1,1)$.
\end{enumerate}
Both restrictions are consequences of the assumption that
there are no tensile forces anywhere along the cell edge.
The maximal couple that can be produced by a given normal force amplitude
corresponds to the case in which the entire normal force 
is applied at one corner of the cell, 
in which case $\aone$ (or $\atwo$) is $\pm 1$.  
The assumption is not rigorously valid since 
there can be tensile forces in the interior of a grain that is
cut by the edge of a cell.
The central hypothesis of the \amodel, however, 
is that the geometric randomness in the grain positions can be replaced
by the randomness in the choice of the solution for each cell.
In this context, restrictions 1 and 2 are valid as long as the
cell size is larger than or equal to the grain size.

It is duly noted that the method of propagating forces down from 
the top of the system does not faithfully represent
the physics in the following sense.
The equations of stress equilibrium are elliptic equations
whose solutions depend on the boundary conditions
on all of the system's boundaries.
For the propagation method employed in the \amodel,
the only cells that can be affected by a change in the
distribution of forces on a given cell are the ones
that lie in a downward opening cone with $45^{\circ}$ edges,
and in this respect the model is more appropriate to a
hyperbolic system.
The goal here, however, is to construct the simplest model
consistent with the relevant stress balance constraints
and capable of displaying nontrivial fluctuations.
In addition, it should be noted that the classical approach
to the computation of stresses via Mohr-Coulomb constitutive
relations also transforms the problem into a hyperbolic one,
and the computational method of propagating stresses downward 
from the top is routinely exploited in this context as well. \cite{pitman}
The development of rigorous elliptic methods for generating 
consistent configurations is an interesting 
and potentially important problem as well, 
but is beyond the scope of this work.

The adoption of a hyperbolic method necessarily leads to 
a third restriction on the force amplitudes:
\begin{list}{}{}
\item[3.] All tangential force amplitudes must be positive. 
\end{list}
In the absence of this restriction it is impossible to ensure
that restriction 1 above can be met for every cell.
That is, if negative tangential amplitudes are permitted to occur,
then it can happen that for some cell deeper in the system
all solutions to 
Eqs.~(\ref{eq:vforcebalance}),~(\ref{eq:hforcebalance}), 
and~(\ref{eq:torquebalance})
have a negative value of either $\none$ or $\ntwo$.
On the other hand, it can be shown, 
using the method described in the following subsection, 
that restriction 3 guarantees a consistent solution for
every cell.

The final assumption defining the \amodel\ is that 
geometric randomness in the structure 
of a granular material can be effectively modeled 
by assigning equal probability to
all possible solutions to the force and torque balance equations 
at each cell.
This requires defining a particular measure on the
space of solutions.
In the absence of any good reason to choose otherwise, 
it will be assumed that all values of the triple $(\azero,\aone,\atwo)$
that are consistent with restrictions 1-3 above
are equally probable.
{\em Throughout the rest of this paper it will also be assumed, for simplicity,
  that all cells have the same $m$ and that $u=0$ for every cell.}

This concludes the definition of the 2D \amodel\ for 
non-cohesive granular materials consisting of convex grains.
The next subsection presents a geometric picture of the solution space
of the force and torque balance equations for a single cell.
The picture provides useful insight into the need for restriction 3
and also reveals that the cell size in the model must be interpreted
as the grain size.

The \qmodel\ may be thought of as a rather drastic simplification of
the \amodel\ in which 
Eqs.~(\ref{eq:hforcebalance}) and~(\ref{eq:torquebalance})
are ignored.
The horizontal components of the forces can then be neglected
and it becomes more convenient to work with variable 
$w_i \equiv (n_i+t_i)/sqrt{2}$ and 
$W_i \equiv (N_i+T_i)/sqrt{2}$.
Eq.~(\ref{eq:hforcebalance}) can then be written as
\begin{eqnarray}
  \label{eq:qmodel1}
  w_1 & = & q (W_1+W_2+mg) \\
  \label{eq:qmodel2}
  w_2 & = & (1-q) (W_1+W_2+mg).
\end{eqnarray}
In the \qmodel, $q$ at each site is an independent random number
chosen from some distribution with support only on the interval $[0,1]$.
The lack of correlation between $q$'s at different sites
is an important feature in the asymptotic analysis of 
Coppersmith et al.\cite{qmodel1}
Note that the \amodel\ differs from the \qmodel\ in that the
random variables at each cell cannot be chosen in advance,
but only after the forces have been propagated down to the cell.
If all of the $\alpha_i$ are fixed in advance,
negative tangential and normal forces are quickly generated
and the amplitudes diverge rapidly with increasing depth.
Thus the ``maximally random'' \amodel\ does contain correlations
that are direct consequences of the additional force and torque
balance constraints.
In this sense it is related to attempts to derive force distributions
by maximizing a statistical entropy associated with the possible
ways of constructing microscopic configurations consistent with
an imposed average stress field \cite{bagiPG}, but differs in that
the average stress field in the \amodel\ is not given in advance.

\subsection{Solving the force and torque balance equations in a single cell} 

For the purposes of this subsection we will take $g=0$.
The effects of gravity can be included in a straightforward manner
once the graphical method is understood. 

All of the information contained in the values of 
$N$, $C$, and $T$ on one edge
can be encoded in a single vector placed at some position along the edge.
The vector itself represents the sum of the normal and tangential forces
and the position is chosen such that the torque that would be produced about
the center of the cell by the normal component of the force is equal to $C$.
Thus the net effect of any configuration of stresses 
in the cell can be expressed
graphically by drawing the four vectors, one on each edge.

\begin{figure}[tbp]
  \centerline{
    \includegraphics[width=2.0in]{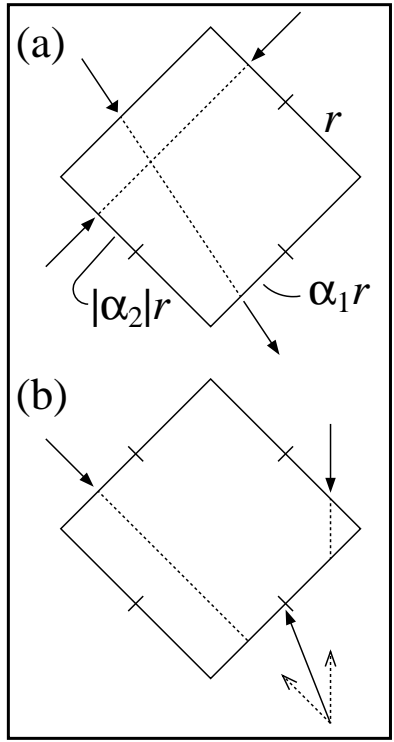}
  }
  \vspace{0.2in}
  \caption{The direct solution for a single cell.
        (a) A straightforward case in which forces are propagated
        directly through the cell.
        (b) A case in which the propagated force rays intersect
        the same edge of the cell.
        The two vector forces must then be summed as shown to obtain
        the solution.}
  \label{fig:directsolution}
\end{figure}

Now given the vectors on the upper edges of a cell 
we would like to determine all the
possible ways of assigning vectors to the lower edges.
One solution is immediately obvious and 
will be called the ``direct solution''.
Simply construct rays originating from the vectors on each edge,
then take the lower vectors to be equal and opposite to the upper vectors and
positioned where their respective rays intersect a lower edge.
(See Figure~\ref{fig:directsolution}.)
Each ray will be called a ``ray of force'', 
though it is understood that the stresses represented are not really
concentrated on the ray.
The only complication that can arise is that both rays of force
intersect the same lower edge.
In this case, 
the vector assigned to this edge is the sum of the two, 
positioned such that the couple associated with it is the 
sum of the couples associated with the two.
(See Figure~\ref{fig:directsolution}.)
Note that the direct solution is guaranteed to exist if and only if 
each ray of force is guaranteed to intersect a lower edge,
rather than exiting the cell through the other upper edge.
This will be true whenever the normal and tangential amplitudes
on the upper edge are both positive and 
is directly related to restriction 3 above.

\begin{figure}[tbp]
  \centerline{
    \includegraphics[width=2.5in]{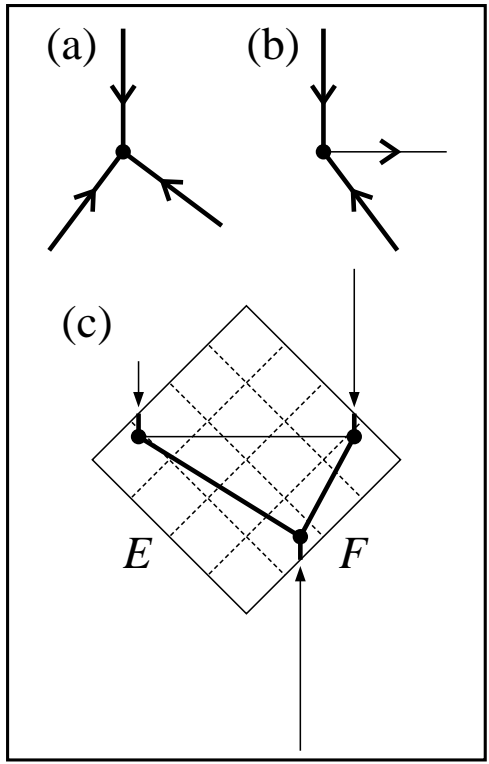}
  }
  \vspace{0.2in}
  \caption{splitting of force rays and tensile forces within a cell.
        (a) A vertex involving only positive force amplitudes.
        (b) A vertex involving a negative force amplitude.
        (c) A solution in a cell involving a negative force amplitude.
        The smaller dashed cells can be used to argue that real tensile
        forces must be associated with this configuration.
        }
  \label{fig:forcerays}
\end{figure}

To construct solutions different from the direct solution, 
note that force and torque balance can be preserved by having
a ray of force split into two rays at some arbitrary point in the cell,
as shown in Figure~\ref{fig:forcerays}a.
The force associated with each ray is directed along the ray
and the force amplitudes are then fixed by requiring that the 
three forces sum to zero.
Torque balance is guaranteed because none of the three forces 
generates any torque about the splitting point.
Similarly, two rays of force that intersect can be merged 
into a single ray of force emanating from the point of intersection.

Whenever all three rays of force intersecting at a single point 
lie in the same half-plane, the force amplitude associated
with one of them must be negative.
This situation is depicted in Figure~\ref{fig:forcerays}b.
The thin line in the figure indicates a negative amplitude.
Any solution of Eqs.~(\ref{eq:vforcebalance})-(\ref{eq:torquebalance})
can be depicted as a network of force rays within the cell.
An example involving a negative amplitude 
is shown in Figure~\ref{fig:forcerays}c.

It is possible for negative amplitudes to be an artifact
of the summing of two forces that occur at different positions on
the same edge, as would arise for the configuration shown in 
Figure~\ref{fig:directsolution}c.
There exist configurations, however, for which the negative
amplitude can only be interpreted as corresponding to a 
real tensile force.
An example is shown in Figure~\ref{fig:forcerays}c.
The forces applied to the upper edges of the cell result in
{\em no} force being transmitted across edge $E$.
The force transmitted across edge $F$ balances both the forces
and net torque produces by the forces above.
There is no way to produce the configuration shown in a material
that does not support tensile forces.
To see this, consider the description of the configuration using cells
of smaller scale (the dashed cells in Figure~\ref{fig:forcerays}c.
It can be shown that it is impossible to construct a network of
force rays on the smaller cells that would be represented by the
network shown on the large cell, without having tensile segments
that cross the boundaries of some of the small cells.
(The situation corresponds to that of a horizontal beam clamped 
at one end and supporting a load on the other, in which case
the top portion of the beam is under tension.)

Thus if all possible solutions of the force and torque balance equations
occur with equal probability, the material within a cell must be assumed
to support tensile forces and hence must be on the order of or smaller than
the grain size.
Together with the reasoning leading to restriction 1,
this leads to the conclusion that 
{\em the cell size in the \amodel\ must be the grain size.}

In practice, the region in $\alpha$-space which yields
solutions that satisfy restrictions 1-3 is not easily identified.
Restriction 2 immediately implies that $\aone$ and $\atwo$ must
lie in the range $(-1,1)$.
Using Eqs.~(\ref{eq:vforcebalancea})-(\ref{eq:torquebalancea2})
and the restrictions 1-3, it can also be shown that
$\azero$ must lie in the range
\begin{equation}
  \label{eq:azerorange1}
 -\None - \Ntwo + \Cone + \Ctwo - \sqrt{2}mg
  \leq
  2 \azero 
  \leq 
  \None + \Ntwo + \Cone + \Ctwo + 2\Tone + 2\Ttwo - \sqrt{2}mg.
\end{equation}
Thus the entire region of consistent solutions must be contained within
a rectangular region in $\alpha$-space.
To give equal weight to every solution consistent with the restrictions,
a point is selected at random according to a uniform probability density
throughout the rectangular box.
The values of $\none$, $\ntwo$, $\tone$, and $\ttwo$ are then computed
from Eqs.~(\ref{eq:vforcebalancea})-(\ref{eq:torquebalancea2}).
If any of these quantities is negative, the solution is discarded
and the process is repeated until a consistent solution is found.
Since every point within the rectangular box has an equal probability
of being chosen on every attempt, 
it follows that every point that corresponds to a consistent solution
has an equal probability of being selected.

In some cases, it may turn out that the set of consistent solutions
occupies a very small region of the rectangular box so that the
probability of finding a solution by random guessing is prohibitively small.
The results described in this paper were obtained by imposing a
cutoff $K$ on the number of unsuccessful guesses.
If the cutoff was reached for a  particular cell, 
the direct solution for that cell was used.

A final issue that must be addressed is how the gravitational
force is to be distributed in the direct solution.
Some care must be taken here to avoid generating more and more
cells for which the direct solution must be used.
The choice made for the simulations described in this paper
is that the contributions due to the weight of the cell are
\begin{eqnarray}
  \none & = & \sqrt{2}mg \mu/(1+\mu)(1+a_1), \\
  \ntwo & = & \sqrt{2}mg /(1+\mu)(1+a_2),    \\
  \cone & = & a_1 \none, \\
  \ctwo & = & a_2 \ntwo, \\
  \tone & = & \cone, \\
  \ttwo & = & \ctwo,
\end{eqnarray}
where $a_1$ and $a_2$ are each random numbers between 0.1 and 0.5, and 
$\mu \equiv (\None + \Ttwo + mg/\sqrt{2})/(\Ntwo + \Tone + mg/\sqrt{2})$.
With this choice, the portion of the weight transmitted to the 
right (or left) increases when the cell is being 
pushed to the right (or left) from above.
This is meant only to be plausible.
It enters the computation of the stresses only when the
direct solution is used, which can be made as infrequent
as desired by raising the value of the cutoff $K$.

\subsection{Boundary conditions}

The two sections below discuss the configurations generated under
different sets of boundary conditions.
In section~\ref{sec:silo}, the focus is on the average stresses in
a deep, narrow silo geometry, where walls can (and do) support stresses.
The model then assumes the geometry shown in Figure~\ref{fig:amodel}a
with the shaded cells assumed to absorb all forces applied
from above.
The force and couple amplitudes on the dashed edges 
are taken to vanish identically.
While there is no attempt to model the elasticity of the walls in
a realistic way, significant differences between the \amodel\ and
the \qmodel\ will arise, leading to some useful insights.

In section~\ref{sec:deepbox}, the goal is to explore the asymptotic
behavior of the stresses in an infinitely wide and deep system.
For this purpose periodic boundary conditions are employed in the
horizontal direction.
The two shaded cells in a layer in Figure~\ref{fig:amodel}a
are identified and all cells are treated equivalently.
The width of the system is always taken large enough so that
the maximal force observed on any cell is small compared to
the total force on its layer.

\section{Average stresses and fluctuations in a narrow silo}\label{sec:silo}

One classical approach to the computation of stresses in 
a granular material to assume that the material
satisfies the Mohr-Coulomb criterion everywhere in space.
Denoting the principal stresses as $\sigma_1$ and $\sigma_2$,
the Mohr-Coulomb criterion in 2D reads
\begin{equation}
  \label{eq:mc}
  \frac{\sigma_1-\sigma_2}{\sigma_1+\sigma_2} = \sin \phi,
\end{equation}
where $\phi$ is the internal friction angle characteristic
of the material.
A typical assumption for the boundary condition at the walls
is $\sigma_{xy} = \tan(\phi_w)\sigma_{xx}$,
where $\phi_w$ is a parameter characterizing the friction
between the material and the wall.

Under these assumptions, the stress equilibrium equations can
be solved numerically. \cite{pitman}
The centerline stress $\sigma_{yy}$ at depth $y$ is well fit by
the solution of Janssen 
(which relies on additional simplifying assumptions)
\begin{eqnarray}
  \label{eq:janssen}
  \bar{\sigma_{yy}} = \frac{R \rho g}{\kappa}\left(1-e^{ay/R}\right),
\end{eqnarray}
with $\kappa \equiv  \tan(\phi_w)(1+\sin\phi)/(1-\sin\phi)$.
The Janssen solution assumes that $\sigma_{yy}$ is independent
of horizontal position, which does not turn out to be true in
the full solution. \cite{nedderman,pitman}
Nevertheless, the deviations, which are of the order of 15\% from
wall to centerline, do not alter the fundamental result that
$\sigma_{yy}/R$ is a function of $y/R$ only.

The behavior of the \qmodel\ in this geometry is quite different.
An ensemble average of configurations generated by the \qmodel\ 
yields a parabolic profile for the vertical forces supported.
At large depths a steady state is attained 
in which the vertical stress near the
wall must be proportional to the radius $R$ of the silo,
since the vertical force transmitted to the walls at each layer 
must equal the weight of the layer on average.
The stress at the centerline, on the other hand,
is proportional to $R^2$.
This behavior is accurately reproduced by an analytic calculation
of the distribution of forces in the \qmodel\ when all
$q$'s are assigned their mean value of $1/2$.
The analytic solution also indicates that the approach to
the asymptotic profile occurs more slowly, with a decay
constant of the order of $1/R^2$ rather than $1/R$.
\cite{pfPG}

Experimental results appear to confirm the scaling expectations 
of the classical analysis.
The recent experiments of Cl\'{e}ment et al., for example, 
show a linear scaling with system width in both 
the asymptotic average vertical stress and the characteristic
depth of the approach to the asymptotic value. \cite{clementPG}

\begin{figure}[tbp]
  \centerline{
    \includegraphics[width=3.0in]{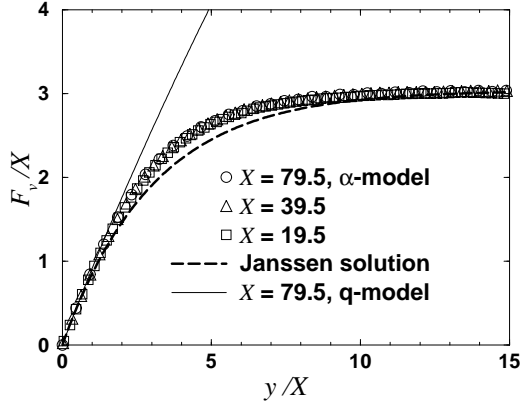}
  }
  \caption{Average vertical force on a layer in the silo geometry.
        The discrete symbols are obtained from simulation of the \amodel.
        (See text for specification of parameters.)
        The thick dashed line is a fit to the Janssen solution, with
        one free parameter chosen to match the asymptotic value of the
        force at large depths.
        The thin line shows the behavior of the \qmodel\ with a 
        uniform distribution of $q$'s.}
  \label{fig:silo}
\end{figure}

Results for the average vertical force as a function of depth in the
\amodel\ are shown in Figure~\ref{fig:silo}.
Figure~\ref{fig:silo} shows data for three different silo widths,
plotted in terms of the scaled variables suggested by the classical analysis.
The average vertical force $F_v$ is here defined as 
the total vertical force transmitted across the corrugated surface 
marked with a thick line in Figure~\ref{fig:amodel}a, 
divided by the number of cells in the surface, $X$:
\begin{equation} \label{eq:Fv}
F_v = \frac{\sum_{\rm layer} 
        \frac{1}{\sqrt{2}}(\none+\tone+\ntwo+\ttwo)}{X}.
\end{equation}
Note that $R = X r \sqrt{2}$.
From Eqs.~(\ref{eq:ndef})-(\ref{eq:tdef}) 
the value of this quantity in the Janssen solution can be calculated.
Accounting properly for the density of the material $\rho = m/(2r)^2$ 
and using the $y = j r \sqrt{2}$, where the integer $j$ indexes the layers,
we find
\begin{equation} \label{eq:FvJanssen}
\frac{F_v}{X} = \frac{1}{\kappa}\left(1 - e^{-\kappa j/X}\right),
\end{equation}
where  $\kappa$ is the combination of 
material and wall parameters defined above. 

Each data point in Figure~\ref{fig:silo} was obtained by averaging $F_v$ 
over 1000 configurations. 
In all cases, the cutoff $K$ was taken to be 1000, resulting
in the direct solution being used for approximately 3\% of the cells.
The data collapse obtained using the scaling suggested by the
Janssen solution is quite good.
The heavy dashed curve in Figure~\ref{fig:silo} is 
the classical prediction of Eq.~(\ref{eq:FvJanssen}),
where the single parameter $a$ has been fit to the asymptotic value
of $F_v$ at large depths.
The same value of $\kappa = 2.15$ was used for all data sets and the
$X$ was taken to be the average number of cells in two successive layers,
not including the wall cells.
It is perhaps worth noting that data from an experiment by Cl\'{e}ment et al.
show the same tendency to lie above the Janssen curve at small depths.
\cite{clementPG}
The solid line in Figure~\ref{fig:silo} is 
the prediction from the simplest version of the \qmodel, in which
the distribution of $q$'s is taken to be uniform over the unit interval,
shown here for $X=39.5$.
The line shown was generated by simulation of the \qmodel, but agrees
perfectly with the analytic solution. \cite{pfPG}
In that solution, $F_v$ scales at large depths like $X^2$ and there
is no value of $\kappa$ that yields 
a satisfactory fit to the Janssen solution.

\begin{figure}[tbp]
  \centerline{
    \includegraphics[width=3.0in]{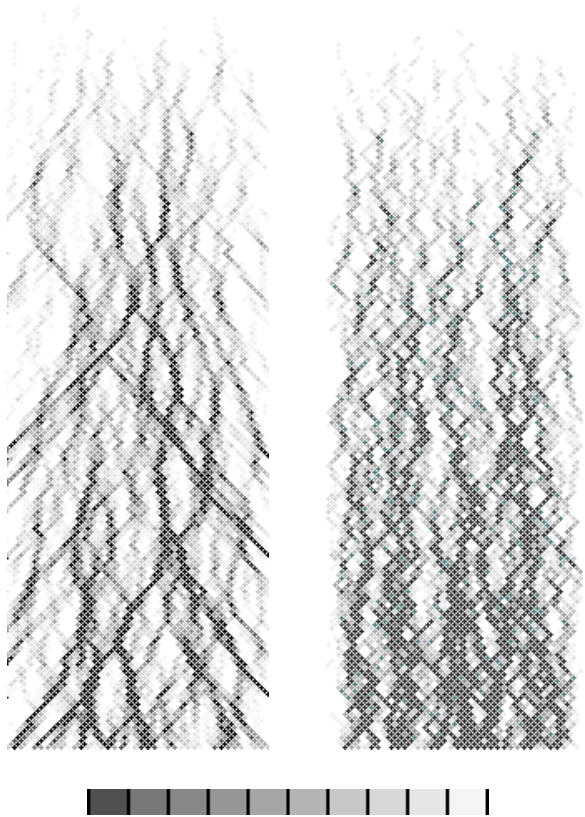}
  }
  \vspace{0.2in}
  \caption{Typical configurations in the silo geometry.
        Darker cells indicate higher values of the vertical force supported.
        The bar indicates how the colors are specified on a linear scale.
        The width of the silo is $X=39.5$.  
        (Excluding the walls, there are 39 cells in each odd layer 
        and 40 in each even one.
        (a) The \amodel. 
        (b) The \qmodel.
        }
  \label{fig:silopic}
\end{figure}

It is clear that the inclusion of the proper force and torque balance
constraints brings the \amodel\ into much closer agreement with
conventional expectations than the \qmodel.
The reason for this appears to be that in the \amodel\ strong stresses 
tend to propagate to the left or right with increasing depth, 
whereas in the \qmodel\ the vertical force simply diffuses.
Figure~\ref{fig:silopic} illustrates this difference with pictures of
vertical force patterns obtained from both models for $X=39.5$.

The \amodel\ clearly exhibits arching on the macroscopic level.
The weight of the material is supported by the walls in the manner
expected from continuum theories.
It is also worth noting that the \amodel\ exhibits well-formed arches
on the ``microscopic'' level, an effect which will be illustrated
more clearly below.
In addition, the filamentary stress chains observed in the \amodel\
are roughly reminiscent of experimental images produced under
a variety of conditions (see, for example, Refs.~\cite{nagel,howellPG}),
and also with numerical solutions of stresses in random disk packings
\cite{radjai}.
Though quantitative comparison with these experiments and models 
is beyond the scope of this work, 
it does appear that the model is capable of displaying
plausible behavior both at the microscopic and macroscopic levels.

Finally, it is instructive to consider the 
\amodel\ in the absence of any randomness.
If $a_1$ and $a_2$ are set to 0.5 and 
the direct solution is employed for every cell,
the model generates a smooth stress field in which the
walls support almost no weight and the vertical force grows
linearly with depth.
This may be thought of as the correct result for a homogeneous solid with
a vanishing Poisson ratio.
The vertical stress generates no horizontal force on the walls,
which therefore cannot bear any weight.
Allowing $a_1$ and $a_2$ to vary between 0.1 and 0.5 but still employing
the direct solution for every cell causes the walls to bear only a
small fraction of the weight of each layer.

This shows that the behavior illustrated in Figure~\ref{fig:silo}
results directly from the randomness in the choice of solutions,
not from the structure of the \nctlattice\ and 
the ``hyperbolic'' method alone. 
No constitutive relation similar or even analogous to
the Mohr-Coulomb condition of incipient yield has been put into
the model, which makes the fact that the results agree reasonably well
with the classical theory quite intriguing.

\section{Fluctuations in a large, periodic box}\label{sec:deepbox}

One of the primary motivations for developing the \amodel\ is to see whether
the predictions of the \qmodel\ survive the inclusion of physically realistic
force and torque balance constraints.
Three questions are of particular importance:
\begin{enumerate}
\item Does the probability distribution for stresses in an 
    infinitely wide system approach a stationary limit at large depths?
\item If so, does the limiting probability distribution 
      have the same form as that of the \qmodel?
\item Do the constraints induce any long range spatial correlations
      in the stresses?
\end{enumerate}
As a first step in answering these questions, 
distributions of the vertical forces
were computed for a system of width 500 at depths up to 450.
The cutoff $K$ was taken to be 1000, 
requiring direct solutions for approximately 3\% of the cells.

\begin{figure}[bp]
  \centerline{
    \includegraphics[width=3.0in]{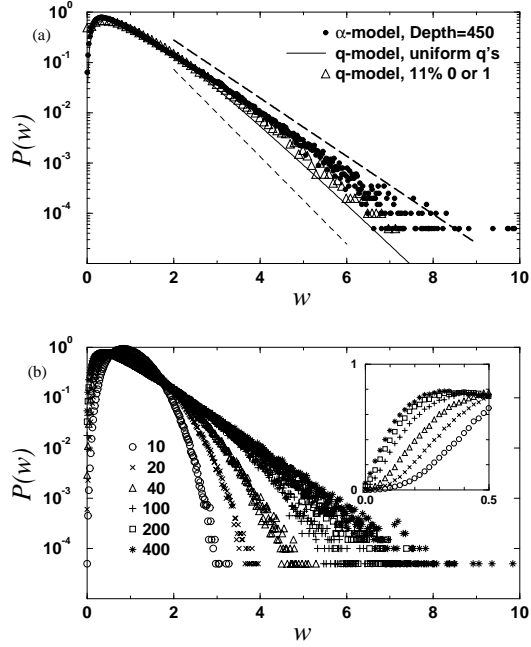}
  }
  \vspace{0.2in}
  \caption{Probability distributions for vertical forces at various depths.
        (a) Comparison of the \amodel\ with $K=1000$ and the 
        \qmodel\ with both a uniform distribution of $q$'s 
        and a distribution
        that is uniform except for sharp spikes at 0 and 1 accounting for
        11\% of the total density.
        The data is from the layer at depth $450$.
        The dashed lines are guides to the eye, with slopes -2 and -1.3.
        (b) The \amodel\ as in (a), but with data shown for 
        several different depths.
        The inset shows the region near $w=0$, where a clear evolution
        of $P(w)$ with depth is observed. 
        }       
  \label{fig:pbcbox}
\end{figure}

Figure~\ref{fig:pbcbox} shows the $P(w)$, the probability density
for the vertical force supported by a single cell at various depths.
The weight $w$ is defined as the actual force supported,
$(\none+\tone+\ntwo+\ttwo)/\sqrt{2})$, divided by the depth of the layer.
Part (a) shows $P(w)$ for the deepest layer measured,
along with the analytic result for the \qmodel\ with a
uniform distribution of $q$'s between 0 and 1 
which decays as $\exp(-2w)$ at large $w$ \cite{qmodel1}.

In the \amodel,
it appears that $P(w)$ decays as $\exp(-\lambda w)$ for large $w$
with $\lambda$ roughly equal to 1.3,
as shown in Figure~\ref{fig:pbcbox}a.
The dashed lines on the figure are guides to the eye, 
having slopes -2 and -1.3.
Figure~\ref{fig:pbcbox}b indicates that
there is a noticeable evolution
of $P(w)$ up to depths as large as 400, with the
decay at large $w$ becoming slower and the number of
cells supporting almost no weight increasing with depth.

The value of $\lambda$ obtained in the \amodel\ can also be obtained
from the \qmodel\ with a suitable choice for the distribution from
which the $q$'s are chosen.
The open triangles in Figure~\ref{fig:pbcbox}a were obtained from
simulation of the \qmodel\ using a distribution of $q$'s that included
delta-function peaks at 0 and 1;
$q$ was taken to be 0 for 5.5\% of the cells, 1 for 5.5\%,
and uniformly distributed between 0 and 1 for the rest.
The percentages were chosen in order to produce $\lambda=1.3$,
as determined by the mean-field calculation described in the appendix.
For comparison, Figure~\ref{fig:qdist} shows the probability density
for the values of $q$ obtained in the \amodel, where $q$ is defined
as the fraction of the vertical component of force on a cell that
is transmitted to one of the cells in the next layer below, 
just as in the \qmodel.
Note that the \amodel\ does appear to generate singularities at
$q=0$ and $q=1$, but these are {\em not} delta functions.
\begin{figure}[tbp]
  \centerline{
    \includegraphics[width=3.0in]{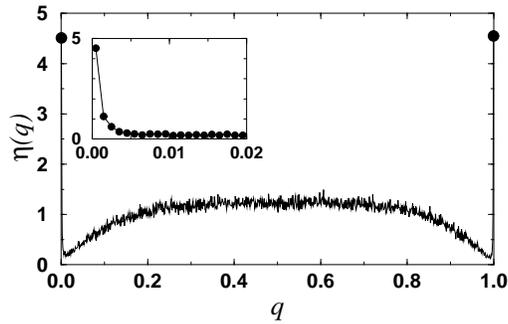}
  }
  \vspace{0.2in}
  \caption{The $q$ distribution generated by the \amodel\ with 
        $K=1000$.  Each data point represents the relative probability
        that the $q$ of a given cell will fall in a bin of width 0.001
        and the plot is normalized to correspond to $\eta(q)$ as defined
        in the appendix.
        The large circles mark the values for the bins centered on 0.0005
        and 0.9995.  The inset shows an expanded view of the singularity
        near $q=0$.
        There is no measurable delta-function contribution
        at $q=0$ or $q=1$.
        }       
  \label{fig:qdist}
\end{figure}

\begin{figure}[tbp]
  \centerline{
    \includegraphics[width=3.0in]{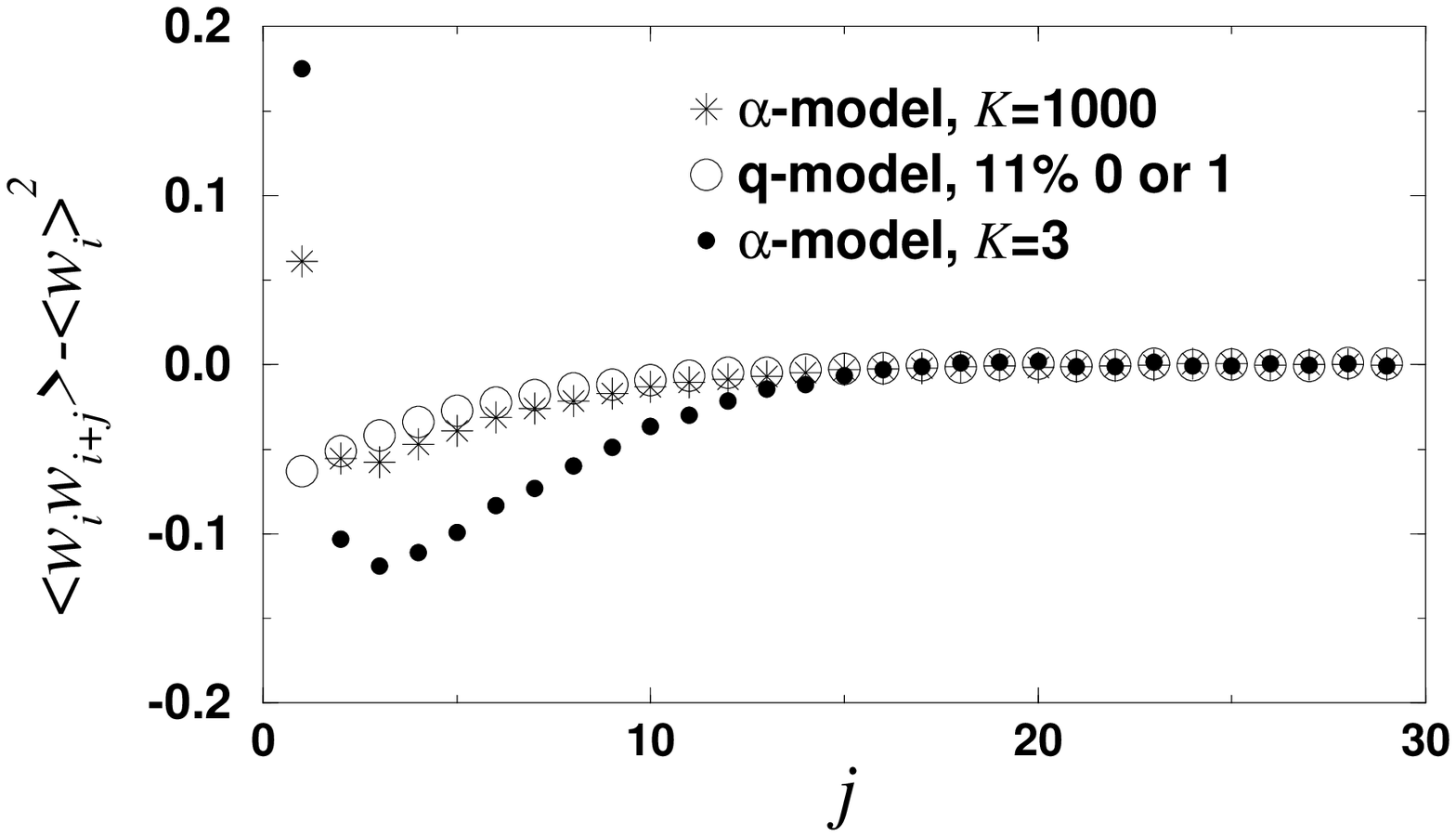}
  }
  \caption{Correlations between weights on cells in one layer.
        The averaging was done over all cells at depth 200
        in 9000 configurations.
        Data is shown for the \amodel\ with $K=1000$ and $K=3$ and the 
        \qmodel\ with a $q$ distribution that is uniform except for 
        delta functions at 0 and 1 accounting for 11\% of the total density.
        }       
  \label{fig:corr}
\end{figure}

Figure~\ref{fig:corr} compares a measure of the spatial correlations in
the \amodel\ and the \qmodel.
The correlation function $\langle w_i w_{i+j}\rangle_c$ is plotted,
where $i$ indexes the horizontal position of a cell in a single layer.
Though there is a discernible difference between the two models,
it is clear that correlations decay rapidly, on the order of 10 cells.
It may be argued that the \qmodel\ predictions should apply to the \amodel\ 
at large $w$ since the force and torque balance constraints do not appear
to generate any long range correlations.
Though there is as yet no analytic proof that the 
\amodel\ weight distribution
will indeed conform to the expected decay rate at very large $w$,
the calculation in the appendix showing that this is the behavior
expected in the \qmodel\ for an appropriately chosen $q$ distribution,
together with the fact that the $q$ distribution obtained from the
\amodel\ is reasonably well approximated by a singular distribution of
this type, strongly suggest the conclusion that the exponential decay
with $\lambda\approx 1.3$ will persist to arbitrarily large $w$.
It must be noted, however, that the numerical data for the \qmodel\ appear
to correspond to a slightly larger value of $\lambda$.
This may be due either to the influence of correlations not taken into
account in the mean field caclulation, or possibly to the fact that
the asymptotic decay rate emerges only for larger $w$ or larger depths 
than were accessed in the simulations of Figure~\ref{fig:pbcbox}
Thus it is difficult to extract a more accurate value of $\lambda$ 
for the \amodel\ from the data available at present.

The maximally randomized \amodel\ produces behavior more closely
approximated by the \qmodel\ with 11\% 0's and 1's than by the maximally
randomized \qmodel.
Indeed, the two models yield rather similar spatial correlations as well
as single site weight distributions.
It is interesting to note that Radjai {\em et al.} reported an exponential
decay in $P(w)$ with $\lambda=1.4$ in numerical solutions for the
stresses in 2D disk packings in 
squares of side length $\sim 30$ disk diameters
subjected to large external loading \cite{radjai},
not far from the value predicted by the \amodel.
In three dimensions, a similar singular $q$ distribution was also found to
agree best with dynamical simulations of spherical grains. \cite{qmodel1}
It is also worth emphasizing that in both the 
\amodel\ and the \qmodel\ with an 
appropriate $q$-distribution, there is significant evolution of $P(w)$ for 
depths up to 450, even for small values of $w$.

\begin{figure}[tbp]
  \centerline{
    \includegraphics[width=3.0in]{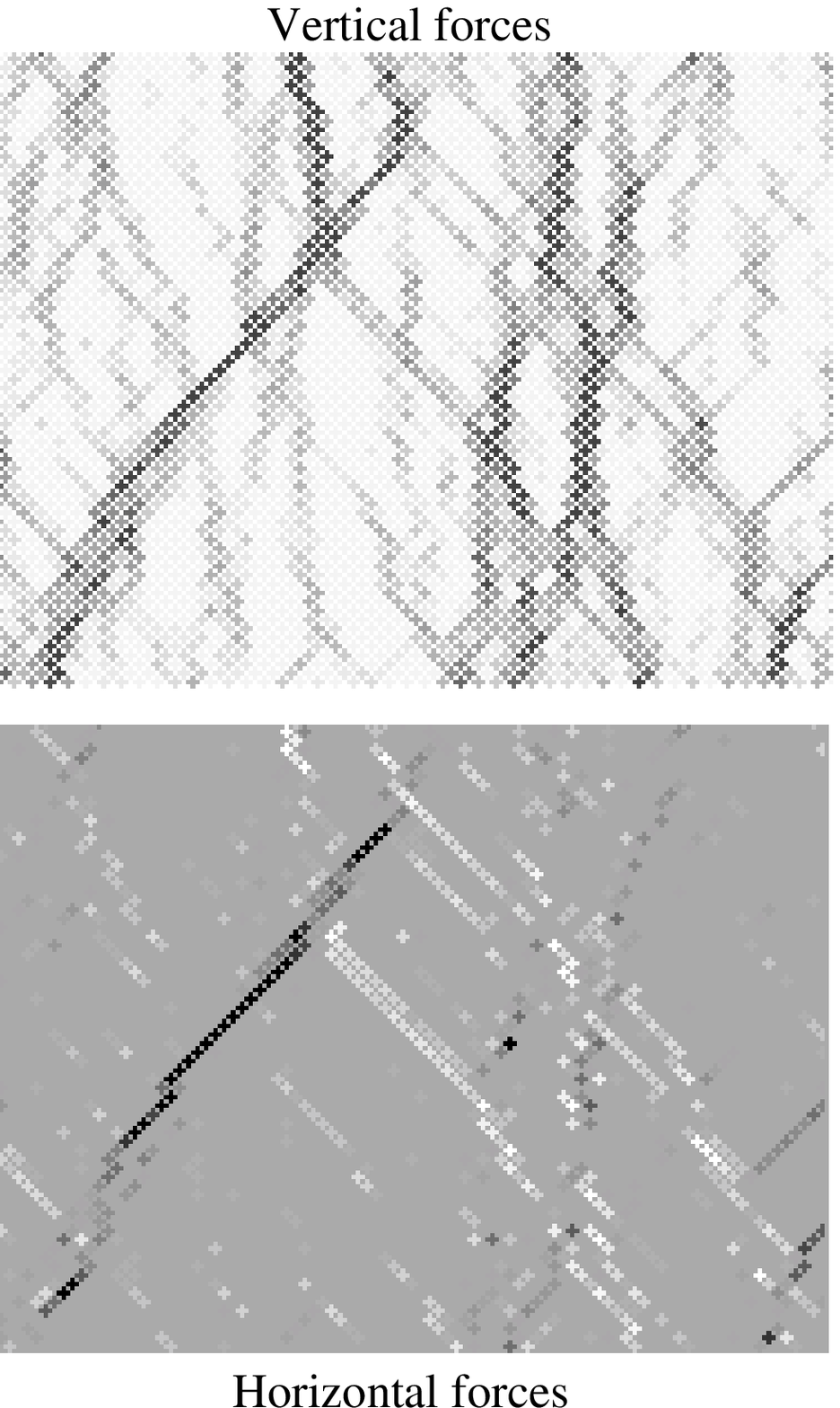}
  }
  \vspace{0.2in}
  \caption{Microscopic arching in the \amodel.
        (a) Vertical forces supported by individual cells in the \amodel\ 
        with $K=1000$.  
        Darker cells support larger vertical forces.
        The picture shown is a detail of a 
        larger configuration, corresponding
        to a section the lower-most 70 layers in a system 240 layers deep.
        A clear central arch can be seen, 
        together with several smaller arches.
        (b) Horizontal forces in the same region as (a).
        Darker cells are being pushed to the left by cells above them
        and lighter ones to the right.
        The arch apparent in (a) is seen here to have the expected structure
        of horizontal forces.
        }       
  \label{fig:pbcboxpic}
\end{figure}

As shown above for the silo geometry, the \amodel\ does provide a
plausible picture of the macroscopic stress field.
Unlike continuum theories, however, the \amodel\ can also provide
details on the scale of the grain size.
Figure~\ref{fig:pbcboxpic} shows a typical portion of a configuration
at large depth for the \amodel.
Both the vertical and horizontal forces applied 
to each cell from above are shown.
These images reveal that weight is supported by a network of arches
with thickness on the order of the grain size.
The appearance of such structures in a random model of this type
is a nontrivial observation, as different ways of choosing
$(\azero,\aone,\atwo)$ can lead to substantially thicker chains
and even nearly uniform distributions.

\begin{figure}[tbp]
  \centerline{
    \includegraphics[width=3.0in]{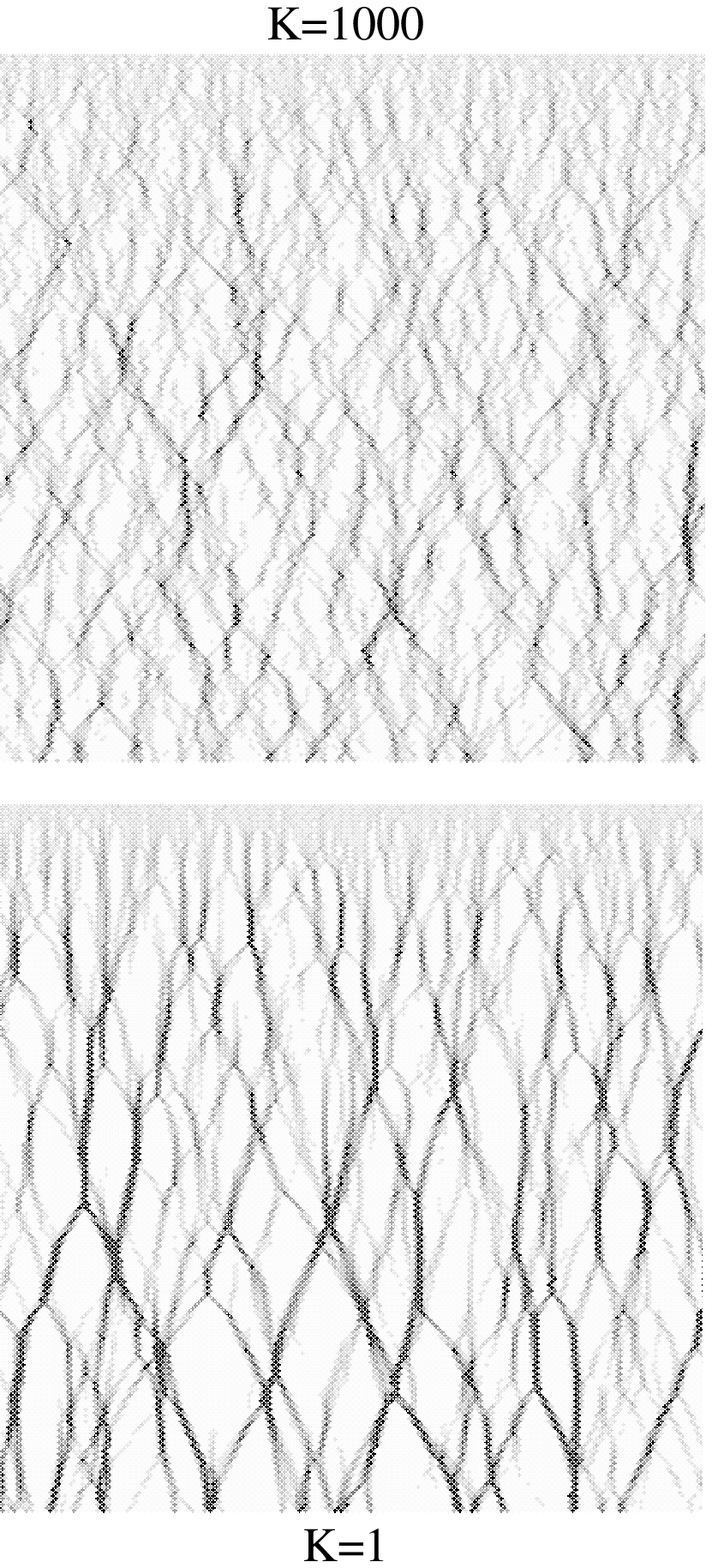}
  }
  \vspace{0.2in}
  \caption{Comparison of the \amodel\ with large and small $K$.
        The vertical force on each cell is shown is for 
        periodic boundary conditions.
        At each layer the force is normalized by the layer depth.
        (a) $K=1000$, resulting in use of the direct solution 
        for 3\% of the cells.
        (b) $K=1$, resulting in use of the direct solution 
        for 80\% of the cells.
        }       
  \label{fig:differentK}
\end{figure}

Finally, a remark on the effect of changing $K$ is in order.
Changing $K$ to 100 in the \amodel\ generates direct solutions at 16\% of
the cells, but has little effect on the results described above.
Changing $K$ to order 1, however, creates a marked increase
in the lengths and directions of the stress chains.
A visual comparison of configurations obtained with $K$ set to 1 or 1000
is shown in Figure~\ref{fig:differentK}.

\section{Conclusions} \label{sec:conclusions}

The \nctlattice\ is
rich enough to describe the stress field in
{\em any} material, with the scale of the cell size being
completely arbitrary.
For the case of non-cohesive granular materials with the cell
size equated with the average grain size, however,
simplifying assumptions can be made that lead to nontrivial predictions.

The \amodel\ studied in this paper includes
particular choices of a few parameters that influence the
details of the distributions.
The latter category of choices has to do with the distribution
of mass within each cell (the parameters $u$, $mg$),
the precise form of the direct solution used 
when random attempts fail,
and the assumption that all force incident on the wall cells
is simply absorbed.
Variations in how these choices are made might be expected
to correspond to different choices for classical parameters
such as the wall-material friction angle 
and the internal friction angle, which would be
reflected in the value of $\kappa$.

Another parameter that can have a noticeable influence on
the force configurations is the cutoff $K$ that roughly determines
how often the direct solution must be used.
When $K$ is small, the direct solution is used often and
the details of how it is implemented can be important.
For large $K$, however, the direct solution is used only
in situations where the range of possible solutions is
highly restricted anyway, so that all possible choices
are actually quite close to the direct solution.
For this reason, the configurations generated with $K=1000$
are excellent approximations to the maximally randomized \amodel.

The present version of the model includes 
an assumption that permits configurations to be
generated by propagating forces down from the top
(restriction 3).
This assumption is {\em not} rigorously justifiable, and
may well be expected to fail in situations
where strain effects are important.
Savage has emphasized the importance of such effects,
especially in the case of a free-standing pile.
The boundary condition at the bottom of the pile,
(the stiffness of the supporting substrate, for example)
is known to be important in determining the stress field. 
\cite{savagePG}
From the perspective taken in the current work, the question
posed by the influence of the boundary conditions is how
the boundary conditions affect the distribution of $\alpha$'s.
Investigation of this issue might be possible if
restriction 3 can be discarded and an algorithm developed for 
finding solutions consistent with appropriate boundary conditions
on all sides of the lattice, including the bottom.
In any case, the \amodel\ is designed primarily to lend 
insight into microscopic and macroscopic fluctuations,
not to investigate the details of how boundary conditions
affect the average stress field.

The solutions obtained from the \amodel\ as
constituted in this paper are sufficiently compatible with
experiments on average stresses \cite{clementPG} and 
fluctuations \cite{qmodel2,miller} to warrant further study. 
The \amodel\ allows study
of the qualitative features of the stresses at the grain size scale
under the simplest physically consistent assumptions for the
form of the geometric disorder.
The effect of the disorder is taken to permit all possible
solutions of the local force and torque balance equations
with uniform probability in the solution space parameterized by
$\azero$, $\aone$, and $\atwo$.
Further work is needed to determine the sensitivity of the results
to changs in the probability measure on this space.

From the data presented in this paper,
it appears that the 2D \amodel\ predicts a weight distribution
that decays as $\exp(-1.3 w)$, consistent with \qmodel\
predictions if and only if an appropriate singular distribution
of $q$'s is used.
For such a $q$ distribution, the \qmodel\ also yields
spatial correlations similar to the \amodel.
This may be taken both as an indication that the primary influence
of the torque and horizontal force balance constraints is to
select a particular form for the probability with which vertical
force is transmitted between adjacent sites, and as a
justification of the use of the \qmodel\ for understanding
the basic structure of the stress fluctuations.

In the silo geometry, the \amodel\ achieves a possibly unexpected
measure of success that is not obtainable by adjusting parameters
in the \qmodel.
The form of the average stresses generated by the \amodel\ agrees 
rather well with experiments \cite{clementPG}.
This type of behavior arises also from the Mohr-Coulomb constitutive
relation, which assumes that the material is at incipient yield everywhere.
The \amodel\ makes the rather different assumption that on the scale of
the grain size the stress is as random as it can be without violating
the fundamental conditions of stress equilibrium. 
The fact that this ``works'' suggests a conceptually new approach
to the description of stress configurations in granular materials.

Another intriguing connection of the \amodel\ to recent work involves
the explicit description of torques at the granular level.
Experimental studies of the thickness of shear bands and also
recent work on continuum models that include the dynamics of
a field that characterizes the local rotation of the material,
known as Cosserat continuum models, have shown that shear bands
tend to have characteristic widths on the order of 10 to 20 
grain diameters. \cite{vardoulakis,tejchman}
The occurrence of a similar length scale in \amodel\ correlations
suggests that the two approaches may be linked in a more profound
way than has yet been understood.

Generalization of the \nctlattice\ and \amodel\ to three dimensions 
is straightforward but requires a 
substantially larger number of variables per cell.
Using a face-centered cubic lattice oriented 
with the $111$ axis on the vertical,
one finds that there are 18 variables that must be computed for each cell.
For each of the three downward-facing faces, one must find
\begin{itemize}
\item[$\bullet$] a normal force,
\item[$\bullet$] two components of the tangential force,
\item[$\bullet$] two components of the couple 
  (about the two axes that lie in the
  plane of the face) associated with the normal force,
\item[$\bullet$] and a third ``torsional'' couple 
(about the axis perpendicular to the face) 
determined by the spatial distribution of the tangential forces.
\end{itemize}
The generalizations of 
Eqs.~(\ref{eq:vforcebalance}),~(\ref{eq:hforcebalance}), 
and~(\ref{eq:torquebalance})
provide six constraints, one for each component of force and torque.
The resulting 12-dimensional space of possible solutions 
can be parameterized
by six $\alpha$'s relating the normal couples to the normal forces,
three more analogous to the $\alpha_0$ of the 2D model,
and the three torsional couples.
The high dimension of the solution space for a single cell makes 
the random guessing approach rather inefficient,
and statistically significant data has not yet been obtained.

\acknowledgements
Conversations with R. Behringer, S. Coppersmith, K. Matveev, R. Palmer,
D. Schaeffer, M. Shearer, and S. Tajima, are gratefully acknowledged.
This work was supported in part by NSF grant DMR-94-12416.

\appendix
\section*{Mean field calculation of $\lambda$ for the \qMODEL}
Coppersmith {\em et al.} have derived several important results concerning
the behavior of $P(w)$ at large $w$ in the \qmodel\ for various choices
of the distribution of $q$'s. \cite{qmodel1}
Define $\eta(q)$ as the probability that a given bond between cells
will carry a fraction $q$ of the vertical force exerted by the higher cell.
Coppersmith {\em et al.} show that $P(w)$ decays as $\exp(-\lambda w)$ 
with $\lambda=2$ in 2D for any $\eta(q)$ that has no singular contributions 
at $q=0$ (or $q=1$).
They also show, however, that different values of $\lambda$ can be obtained
if such singularities are present.

This appendix extends their mean field calculations to the case of 
distributions of the form
\begin{equation}
  \label{eq:etadef}
  \eta(q) = \frac{\theta}{2}\left(\delta(q)+\delta(q-1)\right)+(1-\theta)
\end{equation}
in two dimensions, for which analytic results are obtainable.
The term ``mean field'' here refers to the fact that all correlations
between weights on adjacent sites are neglected.
It is known that the mean field result is exact for certain special
distributions, including the uniform one, and also that even in
cases where it is not exactly correct, the deviations from it are
small for large vertical forces.
The calculation utilizes the Laplace transform formalism described 
in sections IID of Ref.~\cite{qmodel1}.
Several results obtained there will be quoted here 
without explicit derivation.

Let $\tilde{P}(s)$ be the Laplace transform of the steady state 
weight distribution $P(w)$ at large depths.
$\tilde{P}(s)$ satisfies equation 2.11 of Ref.~\cite{qmodel1}, 
reproduced here for the case of two dimensions only:
\begin{equation}
  \label{eq:Pseta}
  \tilde{P}(s)=\left[\int_{0}^{1}\;dq\;\eta(q)\;\tilde{P}(sq)\right]^2 .
\end{equation}
In addition, normalization conditions imply that $\tilde{P}(0)=1$ and 
$\tilde{P}(s)\rightarrow 1-s$ as $s\rightarrow 0$.
Substituting the desired form of $\eta$ yields
\begin{equation}
  \label{eq:Pstheta}
  \tilde{P}(s)=\left[\frac{\theta}{2}(1+\tilde{P}(s))+
    (1-\theta)\int_{0}^{1}\;dq\;\tilde{P}(sq)\right]^2 .
\end{equation}
Defining $\tilde{R}(s)=\sqrt{\tilde{P}}$ 
and changing variables in the integral, we have
\begin{equation}
  \label{eq:Rstheta}
  \tilde{R}(s)=\frac{\theta}{2}(1+\tilde{R}^{2}(s))+
    (1-\theta)\frac{1}{s}\int_{0}^{s}\;dx\;\tilde{R}^{2}(x).
\end{equation}
Multiplying by $s$ and differentiating both sides gives
\begin{equation}
  \label{eq:diffeq}
   \frac{d\tilde{R}}{ds} = \frac{\frac{\theta}{2}-\tilde{R}+
         \left(1-\frac{\theta}{2}\right)\tilde{R}^2}{s\;
                                               (1-\theta \tilde{R})},
\end{equation}
which can be solved for $s$ in terms of $\tilde{R}$, yielding
\begin{equation}
  \label{eq:sofR}
  b\;s = \frac{\tilde{R}-1}{(\theta + (\theta-2)\tilde{R})^{
                                                  (2+\theta)/(2-\theta)}},
\end{equation}
where $b$ is a constant of integration.
Note that the normalization conditions on $\tilde{P}$ imply 
$\tilde{R} = 1 - s/2$ in the vicinity of $s=0$.
Substituting this form for $\tilde{R}$ in Eq.~(\ref{eq:sofR}), 
expanding about $s=0$ on the right hand side, 
and equating coefficients of the first order term yields
\begin{equation}
  \label{eq:b}
  b = \frac{1}{2}(2\theta-2)^{-(2+\theta)/(2-\theta)}.
\end{equation}

The inverse Laplace transform of $\tilde{P}(s)$ will be proportional to
$\exp(s_0 w)$ at large $w$, with $s_0$  being the largest value of $s$
for which $\tilde{P}(s)$ has a singularity, 
which occurs wherever $\tilde{R}(s)$
either has a pole or is of the form $c+(s-s_0)^{\nu}$ 
with the constant $c\neq 0$
and the exponent $\nu$ being non-integral.
(If $c=0$, then half-integral $\nu$ also 
does not yield a singularity in $\tilde{R}^2$.)
Although Eq.~(\ref{eq:sofR}) cannot be inverted to 
obtain $\tilde{R}(s)$ explicitly,
the position of the singularity in $\tilde{R}(s)$ can be determined.
First note that for $0<\theta<1$ Eq.~(\ref{eq:sofR}) implies
that $\tilde{R}(s)$ can diverge only at $s=0$;
since the coefficient of $\tilde{R}$ in the denominator 
has magnitude greater than unity and
the exponent is greater than one, the denominator 
must grow in magnitude faster
than the numerator as $|\tilde{R}|$ goes to infinity.
The possible divergence at $s=0$ arises only because we multiplied by $s$
to obtain Eq.~(\ref{eq:diffeq}).
(As mentioned above, $\tilde{R}$ is known to approach 1 at $s=0$.)
Next note that $d\tilde{R}/ds$ {\em must} diverge at any value of $s$
for which $\tilde{R}=1/\theta$, and only at those points, 
as is evident from Eq.~(\ref{eq:diffeq}) 
and the fact that $\tilde{R}$ itself does not diverge.
Finally, repeated differentiation of Eq.~(\ref{eq:diffeq}) reveals that
higher derivatives of $\tilde{R}$ cannot diverge at any point where
the first derivative does not diverge.
Thus the singularity in $\tilde{R}$ 
can be located by setting $\tilde{R}=1/\theta$
in Eq.~(\ref{eq:sofR}) and combining with Eq.~(\ref{eq:b}) 
to determine $s$.
The result is that the singularity occurs at
\begin{equation}
  \label{eq:s0}
  s_0 = 2\left(1-\frac{1}{\theta}\right)
         \left(\frac{2\theta}{2+\theta}\right)^\frac{2+\theta}{2-\theta}.
\end{equation}
$s_0$ approaches 0 as $\theta$ approaches 0, 
which may be expected given that $\theta=0$ corresponds to the critical 
distribution for which the \qmodel\ exhibits 
power law decay in $P(w)$. \cite{qmodel1}
Moreover, $s_0$ approaches -2 as $\theta$ approaches 1, indicating a smooth
convergence to the result derived for the uniform distribution.
(The case of $\theta=0$ must be treated separately, however, 
and it is seen that $\tilde{R}$ develops a pole at -2.)
Finally, the exponent $\nu$ at the singularity may be obtained from
$\lim_{\tilde{R}\rightarrow 1/\theta}\ln(\tilde{d^2 R/ds^2}/\ln(dR/ds)
 =(\nu-2)/(\nu-1)$, which yields $\nu=1/2$.
Thus there is a single singularity in $\tilde{R}(s)$ and in the vicinity
of the singularity we have $\tilde{R} = (1/\theta) + \sqrt{(s-s_0)}$,
with $s_0$ given above.
This result is consistent with the claim in Ref.~\cite{qmodel1} that
$\tilde{P}(s)$ has a square root singularity for any $\eta(q)$ having
a delta-function component at $q=0$.

In order to compare to the numerical results for the \amodel,
it is useful to find the value of $\theta$ that produces a decay
with $\lambda=1.3$.
From Eq.~(\ref{eq:s0}) one sees that $s_0 \approx -1.30\ldots$ 
is produced by $\theta= 0.11$,
which is the reason that this value was chosen for plotting
in Figures~6 and~8.

\end{document}